\def\year{2022}\relax
\documentclass[letterpaper]{article} 
\usepackage{aaai22}  
\usepackage{times}  
\usepackage{helvet}  
\usepackage{courier}  
\usepackage[hyphens]{url}  
\usepackage{graphicx} 
\usepackage{natbib}  
\usepackage{caption} 
\DeclareCaptionStyle{ruled}{labelfont=normalfont,labelsep=colon,strut=off} 
\usepackage{newfloat}
\usepackage{listings}
\usepackage{booktabs}       
\usepackage{amsfonts}       
\usepackage{nicefrac}       
\usepackage{microtype}      
\usepackage{xcolor} 
\usepackage{algorithmic}
\usepackage{algorithm}
\usepackage{color}
\usepackage{xspace}
\usepackage{microtype}
\usepackage{graphicx}
\usepackage{nicefrac}
\usepackage{booktabs} 
\usepackage{amsmath}
\usepackage{Definitions}
\usepackage{subfig}
\usepackage{stackengine}
\usepackage{wrapfig}
\usepackage{multirow}
\newcommand{\pend}{{}}
\newcommand{\kernel}{\kappa}
\floatstyle{ruled}
\newfloat{listing}{tb}{lst}{}
\floatname{listing}{Listing}

\newcommand{\Qr}{\mathcal{Q}}
\newcommand{\Cr}{\mathcal{C}} 
\newcommand{\unw}{U}
\newcommand{\xhdr}[1]{\vspace{1mm} \noindent{{\bf #1.}}}
\newcommand{\para}[1]{\vspace{1mm} \noindent{{#1}}}

\newcommand{\given}{\,|\,}
\newcommand{\our}{{\textsc{NeuroSeqRet}}\xspace}
\newcommand{\querytt}{\texttt{query}}
\newcommand{\Querytt}{\texttt{Query}}

\newcommand{\keytt}{\texttt{key}}
\newcommand{\Keytt}{\texttt{Key}}

\newcommand{\valuett}{\texttt{value}}
\newcommand{\Valuett}{\texttt{Value}}
\newcommand{\bhbc}{\overline{\bm{h}} ^{(c,q)}}

\newcommand{\ourf}{\textsc{SelfAttn-}\our}
\newcommand{\ours}{\textsc{CrossAttn-}\our}

\newcommand{\cp}{\backslash}
\newcommand{\tpprank}{\texttt{Rank}}
\makeatletter
\g@addto@macro{\normalsize}{%
\setlength{\abovedisplayskip}{2pt plus1pt}%
\setlength{\abovedisplayshortskip}{2pt plus1pt}%
\setlength{\belowdisplayskip}{2pt plus1pt}%
\setlength{\belowdisplayshortskip}{2pt plus1pt}}
\makeatother
\usepackage{enumitem}
\newcommand{\bracex}[1]{\left(#1\right)}
\newcommand{\newq}{{q{'}}}

\RequirePackage[textsize=scriptsize,color=yellow!25,linecolor=brown]{todonotes}
\newcommand{\viz}{\emph{viz.}}

\newcommand{\Ucal}{\mathcal{U}}
\usepackage{paralist}

\newcommand{\hash}{\bm{\zeta}}
\newcommand{\ehdr}[1]{\vspace{1mm}\noindent\emph{---#1}}
\newcommand{\intensity}{\rho}

\renewcommand{\etc}{\emph{etc}}

\newcommand{\eat}[1]{}

\title{Learning Temporal Point Processes for\\ Efficient Retrieval of Continuous Time Event Sequences}

\author{Vinayak Gupta$^*$, Srikanta Bedathur$^{*}$ and Abir De$^{\dagger}$}
\affiliations{$^*$ IIT Delhi, $^{\dagger}$ IIT Bombay\\ \{vinayak.gupta, srikanta\}@cse.iitd.ac.in, abir@cse.iitb.ac.in}
\begin{document}

\maketitle

\begin{abstract}
 Recent developments in predictive modeling using marked temporal point processes (MTPP) have enabled an accurate characterization of several real-world applications involving continuous-time event sequences (CTESs). However, the retrieval problem of such sequences remains largely unaddressed in literature. To tackle this, we propose \our which learns to retrieve and rank a relevant set of continuous-time event sequences for a given query sequence, from a large corpus of sequences. More specifically, \our first applies a trainable unwarping function on the query sequence, which makes it comparable with corpus sequences, especially when a relevant query-corpus pair has individually different attributes. Next, it feeds the unwarped query sequence and the corpus sequence into MTPP guided neural relevance models.
 We develop two variants of the relevance model which offer a tradeoff between accuracy and efficiency. 
We also propose an optimization framework to learn binary sequence embeddings from the relevance scores, suitable for the locality-sensitive hashing leading to a significant speedup in returning top-K results for a given query sequence. Our experiments with several datasets show the significant accuracy boost of \our beyond several baselines, as well as the efficacy of our hashing mechanism.

\end{abstract}

\section{Introduction} 
The recent developments in marked temporal point processes (MTPP) has dramatically
improved the predictive analytics in several real world applications--- from information diffusion in social networks to healthcare--- by characterizing them with  continuous-time event sequences (CTESs)~\cite{tabibian2019enhancing,gupta2021learning,samanta2017lmpp,de2018demarcating,Valera2014,rizoiu2017expecting,wang2017human,daley2007introduction, initiator,du2015dirichlet,tabibian2019enhancing,srijan,de2016learning,zhang2021learning,du2016recurrent,farajtabar2017fake,jing2017neural,saha2018crpp, gupta2021reformd, likhyani2020colab}.
In this context, given a query sequence, retrieval of \emph{relevant} CTESs from a corpus of sequences 
is a challenging problem having a wide variety of search-based applications.
For example, in audio or music retrieval, one may like to search sequences having different audio or music signatures;
in social network, retrieval of trajectories of information diffusion, relevant to a given
trajectory can assist in viral marketing, fake news detection, \etc.
Despite having a rich literature on searching similar time-series~\cite{blondel2021differentiable,gogolou2020data,alaee2020matrix,yoon2019time,cai2019dtwnet,shen2018accelerating,cuturi2017soft,paparrizos2015k},
the problem of designing retrieval models specifically for CTES has largely been unaddressed in the past. Moreover, as shown in our experiments,
the existing search methods for time sequences are largely ineffective for a CTES retrieval task, since 
the underlying characterization of the sequences vary across these two domains.
%
\subsection{Present Work}
%
In this paper, we first introduce \our, a family of supervised retrieval models for continuous-time event sequences and
then develop a trainable locality sensitive hashing (LSH) based method for efficient retrieval over very large datasets. Specifically, our contributions are as follows:


\xhdr{Query unwarping} 
The notion of relevance between two sequences varies across applications. A relevant sequence pair can share very different individual attributes, which can 
mislead the retrieval model if the sequences are compared as-it-is. In other words, an observed sequence may be a warped transformation of a hidden sequence~\cite{ido,gervini2004self}. To tackle this problem, \our\ first applies a trainable unwarping function on the query sequence before the computation of a relevance score. 
Such an unwarping function is a monotone transformation, which ensures that the chronological order of events across the observed and the unwarped sequences remains the same~\cite{ido}.

\xhdr{Neural relevance scoring model}
%
In principle, the relevance score between two sequences depends on their latent similarity.
We measure such similarity by comparing the generative distribution between the query-corpus sequence pairs.
In detail, we feed the unwarped query sequence and the corpus sequence into a neural MTPP based relevance scoring model, 
which computes the relevance score using a Fisher kernel~\cite{fisher} between the corpus and the unwarped query sequences.
Such a kernel offers two key benefits over other distribution similarity measures, \eg, KL divergence or Wasserstein distance: (i) it computes a natural similarity score between query-corpus sequence pairs in terms of the underlying generative distributions; and, (ii) it computes a dot product between the gradients of log-likelihoods of the sequence pairs, which makes it compatible with locality-sensitive hashing for certain design choices and facilitates efficient retrieval.
In this context, we provide two MTPP models, leading to two variants of \our,
which allows a nice tradeoff between accuracy and efficiency.
%


\noindent{\underline{\ourf}}: {Here, we use transformer Hawkes process~\cite{zuo2020transformer} which computes the likelihood of corpus sequences independently of the query sequence.} Such a design admits precomputable corpus likelihoods, which in turn allows for prior indexing of the corpus sequences before observing the unseen queries. This setup enables us to apply LSH for efficient retrieval.

\noindent{\underline{\ours}}: Here, we propose a novel cross attention based neural MTPP model to compute the sequence likelihoods.
Such a cross-attention mechanism renders the likelihood of corpus sequence dependent on the query sequence, making it a more powerful retrieval model. 
{While \ours is not directly compatible with such a hashing based retrieval, it can be employed in a telescopic manner--- where a smaller set of relevant candidate set is first retrieved using LSH applied on top of \ourf, and then reranked using \ours.}

\noindent Having computed the relevance scores, we learn the unwarping function and the MTPP model by minimizing a pairwise ranking loss, based on the ground truth relevance labels.

\xhdr{Scalable retrieval}
Next, we use the predictions made by \ourf\ to develop a novel hashing method that enables efficient sequence retrieval. More specifically, we propose an optimization framework that compresses the learned sequence embeddings into binary hash vectors, while simultaneously limiting the loss due to compression.
Then, we use locality-sensitive hashing~\cite{GionisIM1999hash} to bucketize the sequences into hash buckets, so that sequences with similar hash representations share the same bucket. Finally, given a query sequence, we consider computing relevance scores only with the sequences within its bucket. Such a hashing mechanism combined with high-quality sequence embeddings achieves fast sequence retrieval with no significant loss in performance.

Finally, our experiments with real-world datasets from different domains show that both variants of \our\ outperform several baselines including
the methods for continuous-time series retrieval. Moreover, we observe that our hashing method applied on \ourf\ can make a trade-off between the retrieval accuracy and efficiency more effectively than baselines based on random hyperplanes as well as exhaustive enumeration.


\section{Preliminaries}
In this section, we first sketch an outline of marked temporal point processes (MTPP) and then
setup our problem of retrieving a set of continuous time event sequences relevant to a given query sequence. 

\subsection{Marked Temporal Point Processes: Overview}

\xhdr{Continuous time event sequences (CTESs)}
Marked temporal point processes (MTPP) are stochastic processes
which capture the generative mechanism of a sequence of discrete events
localized in continuous time. Here, an event $e$ is realized using
a tuple $(t,x)$, where $t\in\RR_+$ and $x\in \mathcal{X}$ are the arrival time and the mark
of the event $e$. Then, we use $\Hcal(t)$ to denote a continuous time event sequence (CTES) where each event has arrived until and excluding time $t$, \ie, $\Hcal(t):=\set{e_i=(t_i,x_i)\given t_{i-1}<t_{i}<t}$. Moreover we use $\Tcal(t)$ and $\Mcal(t)$
to denote the sequence of arrival times $\set{t_i\given e_i\in\Hcal(t)}$  and the marks $\set{x_i\given e_i\in\Hcal(t)}$. 
Finally, we denote the counting process $N(t)$ 
as counts of the number of events happened until and excluding time $t$,
encapsulating the generative mechanism of the arrival times.

\xhdr{Generative model for CTES}  
The underlying MTPP model consists of two components--- (i) the dynamics of the arrival times and (ii) the dynamics of the distribution of marks. 
Most existing works~\cite{du2016recurrent, zhang2019self,mei_icml,MeiE16,shelton,zuo2020transformer} model the first component using an intensity function which explicitly models
the likelihood of an event in the infinitesimal time window $[t,t+dt)$, \ie,  $\lambda^{\pend}(t )= \text{Pr} (dN(t)=1|\Hcal(t))$. In contrast, we use an intensity free approach following the proposal by~\citet{shchur2019intensity}, where
we explicitly model the distribution of the arrival time $t$ of the next event $e$. Specifically,
we denote the density $\intensity$
of the arrival time and  the distribution $m^{\pend}$ of the mark  of the next event 
as follows:
\begin{align}\label{eq:qm}
 \intensity(t) dt & = \text{Pr} (e \text{ in } [t,t+dt) \given \Hcal(t)), \\
 m^{\pend}(x)&=\text{Pr} (x\given \Hcal(t))
\end{align}
As discussed by~\citet{shchur2019intensity}, such an intensity free MTPP model
enjoys several benefits over the intensity based counterparts, in terms of
facilitating efficient training, scalable prediction, computation of expected arrival times, etc.
Given a sequence of observed events $\Hcal(T)$ collected during the time interval $(0,T]$,
the likelihood function is given by:
\begin{align}
 p (\Hcal(T)) =\textstyle  \prod_{e_i=(t_i,x_i)\in\Hcal(T)}  \intensity(t_i) \times   m^{\pend}(x_i)
\end{align}
\subsection{Problem Setup}
Next, we setup our problem of retrieving a ranked list of sequence from a corpus of continuous time event sequences (CTESs)
which are relevant to a given query CTES. 

\xhdr{Query and corpus sequences, relevance labels} We operate on a large corpus of sequences $\set{\Hcal_c(T_c)\given c\in \Cr}$, where $\Hcal_c(T_c)=\{(t^{(c)} _i, x^{(c)} _i)  \given t^{(c)} _i < T_c \}$. We are given a set of query sequences $\set{\Hcal_q(T_q) \given q\in\Qr}$ with $\Hcal_q(T_q)=\{(t^{(q)} _i, x^{(q)} _i) \given t^{(q)} _i < T_q\}$, as well as a query-specific relevance label for the set of corpus sequences. That is, for a given query sequence $\Hcal_q$, we have: 
$y(\Hcal_q,\Hcal_c)=+1$  if $\Hcal_c$ is marked as relevant to $\Hcal_q$ and $y(\Hcal_q,\Hcal_c)=-1$ otherwise. 

We define
$\Cr_{q\rel} =\{c\in\Cr \given y(\Hcal_q,\Hcal_c)=+1  \}, \text{and, }
\Cr_{q\nrel} =\{c\in\Cr \given y(\Hcal_q,\Hcal_c)=-1  \}$, with $\Cr = \Cr_{q\rel}  \cup \Cr_{q\nrel} $. Finally, we denote $T=\max\{T_q,T_c\given q\in\Qr,c\in\Cr\}$ as the maximum time of the data collection.

\xhdr{Our goal} We aim to design an efficient CTES retrieval system, which would return a list of sequences from a known corpus of sequences, relevant to a given query sequence $\Hcal_q$. Therefore, we can view a sequence
retrieval task as an instance of ranking problem. Similar to other information
retrieval algorithms, a CTES retrieval  algorithm  
first computes the estimated relevance $s(\Hcal_q,\Hcal_c)$ of the corpus sequence $\Hcal_c$ for a given query sequence $\Hcal_q$ and then provides a ranking 
of $\Cr$ in the decreasing order of their scores.

\section{\our Model} 
In this section, we describe \our family of MTPP-based models that we propose for the retrieval of continuous time event sequences (CTES). We begin with an outline of its two key components.

%
%
%
 
\subsection{Components of \our}
\label{subsec:components}
\our\  models the relevance scoring function between query and corpus
sequence pairs. 
However the relevance of a corpus sequence to the query is latent and varies widely across applications. 
To accurately characterize this relevance measure, \our\ works in two steps.
First, it unwarps the query sequences to make them compatible for comparing with the corpus sequences. Then,
it computes the pairwise relevance score between the query and corpus sequences using neural MTPP models.

\xhdr{Unwarping query sequences} Direct comparison between a query and a corpus sequence can provide misleading outcomes, since
they also contain their own individual idiosyncratic factors in addition to sharing some common attributes.
In fact, a corpus sequence can be highly relevant to the query, despite greatly varying in timescale, initial time, etc.
In other words, {it may have been generated by applying a warping transformation on a latent sequence~\cite{ido,gervini2004self}}. 
Thus, a direct comparison between a relevant sequence pair may give poor relevance score. 

To address this challenge, we first apply a trainable unwarping function~\cite{ido} $U(\cdot)$ on the arrival times of a query sequence,
which enhances its compatibility for comparing it with the corpus sequences. More specifically, we define $\unw (\Hcal_q):=\{(\unw(t^{(q)} _i), x^{(q)} _i) \}$.
In general, $\unw$  satisfies two properties~\cite{ido,gervini2004self}\eat{ which are}: \emph{unbiasedness}, \ie, having a small value of $\left\|\unw(t)-t\right\|$ and
\emph{monotonicity}, \ie, $ {d\, \unw(t  ) }/{dt} \ge 0$. These properties ensure that the chronological order of the events across both the warped observed sequence and the unwarped sequence remains same.
Such a sequence transformation learns to capture the similarity 
between two sequences, even if it is not apparent due to different 
individual factors, as we shall later in our experiments (Figure~\ref{fig:UU}). 

\xhdr{Computation of relevance scores}
Given a query sequence $\Hcal_q$ and a corpus sequence $\Hcal_c$, we compute the relevance score $s(\Hcal_q,\Hcal_c)$ 
using two similarity scores, \eg, 
(i) a \emph{model independent} sequence similarity score and (ii) a \emph{model based} sequence similarity score.

\ehdr{Model independent similarity score:} 
Computation of model independent similarity score between two sequences is widely studied in literature~\cite{xiao2017wasserstein,mueen2016extracting,su2020survey,dtw_paparrizos,muller2007}. They are computed using different distance measures between two sequences, \eg, DTW, Wasserstein distance, etc. and therefore, can be immediately derived from data without using the underlying MTPP model.
%
%
%
In this work, we compute the model independent similarity score, $\textsc{Sim}_{\unw}(\Hcal_q,\Hcal_c)$, between $\Hcal_q$ and $\Hcal_c$ as follows:
\begin{align}
\textsc{Sim}_{\unw}(\Hcal_q,\Hcal_c)& = -\Delta_{t}(\unw(\Hcal_q),\Hcal_c)-\Delta_{x}(\Hcal_q,\Hcal_c)
\end{align}
where, $\Delta_{t}$ and $\Delta_{x}$ are defined as:
\vspace{-1mm}
\begin{align*}
\Delta_{t}(\unw(\Hcal_q),\Hcal_c)& = \sum_{i=0}^ {H_{\min}} \left| \unw(t^{(q)} _i)- t^{(c)} _i\right| + \hspace{-2mm}\sum_{\substack{t_i \in \Hcal_c\cup\Hcal_q\\ i > |H_{\min}| }}  \hspace{-2mm} (T -t  _i),   \\[-2ex]
\Delta_{x}(\Hcal_q,\Hcal_c)& = \sum_{i=0}^{H_{\min}} \II[x^{(q)} _i \neq x^{(c)}_i] + \big||\Hcal_c|-|\Hcal_q|\big|.
\end{align*}
Here, $H_{\min} = \min\{|\Hcal_q|,|\Hcal_c|\}$, $T=\max\{T_q,T_c\}$ where the events of $\Hcal_q$ and $\Hcal_c$ are gathered until time $T_q$ and $T_c$ respectively; 
$\Delta_{t}(\unw(\Hcal_q),\Hcal_c)$ is the Wasserstein distance between the unwarped arrival time sequence $\unw(\Hcal_q)$ and the corpus sequence~\cite{xiao2017wasserstein} and,
$\Delta_{x}(\Hcal_q,\Hcal_c)$ measures the matching error for the marks, 
wherein the last term penalizes the marks of last $|\Hcal_c|-|\Hcal_q|$ events of $|\Hcal_c|$.

\ehdr{Model based similarity score using Fisher kernel:}
We hypothesize that the relevance score $s(\Hcal_q,\Hcal_c)$ also depends on a latent similarity which may not be immediately evident from the observed query and corpus sequences even after unwarping. 
Such a similarity can be measured by comparing the generative distributions of the query-corpus sequence pairs.
To this end, we first develop an MTPP based generative model $p_{\theta}(\Hcal)$ parameterized by $\theta$ and then
compute a similarity score using the Fisher similarity kernel between the unwarped query and corpus sequence pairs $(\unw(\Hcal_q),\Hcal_c)$~\cite{fisher}.
Specifically, we compute the relevance score between the unwarped query
sequence $\unw(\Hcal_q)$ and the corpus sequence $\Hcal_c$ as follows:
\begin{align}
\hspace{-2mm} \kernel_{p_{\theta}}(\Hcal_q,\Hcal_c) = \vb_{p_{\theta}}(\unw(\Hcal_q)) ^\top \vb_{p_{\theta}}(\Hcal_c)   \label{eq:fisher}
\end{align}
where $\theta$ is the set of trainable parameters; $\vb_{p_{\theta}} (\cdot)$ is given by
\\[-2ex]
\begin{align}
 \vb_{p}(\Hcal) =  \Ib^{-1/2} _{\theta} \nabla_{\theta} \log p_{\theta}(\Hcal)/||\Ib^{-1/2} _{\theta} \nabla_{\theta} \log p_{\theta}(\Hcal)||_2,\nn
\end{align}
$\Ib_{\theta}$ is the Fisher information matrix~\cite{fisher}, \ie,
$\Ib_{\theta} = \EE_{\Hcal\sim p_{\theta}(\bullet)}\left[\nabla_{\theta} \log p_{\theta}(\Hcal)\nabla_{\theta} \log p_{\theta}(\Hcal) ^\top\right]$.
We would like to highlight that $ \kernel_{p_{\theta}}(\Hcal_q,\Hcal_c)$ in Eq.~\eqref{eq:fisher} is a normalized version of Fisher kernel since $||\vb_{p_{\theta}}(\cdot)||=1$. Thus, $\kernel_{p_{\theta}}(\Hcal_q,\Hcal_c)$ measures the cosine similarity between $ \vb_{p_{\theta}}(\unw(\Hcal_q))$ and $\vb_{p_{\theta}}(\Hcal_c) $.

Note that, KL divergence or Wasserstein distance could also serve our purpose of  computing the latent similarity between  the generative distributions.
However, we choose Fisher similarity kernel because of two reasons: (i) it is known to be a natural similarity measure which allows us
to use the underlying generative model in a discriminative learning task~\cite{fisher,sewell2011fisher}; and, (ii) unlike KL divergence or other distribution
(dis)similarities, it computes the cosine similarity between $ \vb_{p_{\theta}}(\unw(\Hcal_q))$ and $\vb_{p_{\theta}}(\Hcal_c) $, which makes it compatible with locality sensitive hashing~\cite{charikar2002similarity}.

Finally, we compute the relevance score 
as
\begin{align}\label{eq:relevance-score-function}
s_{p,\unw}\left(\Hcal_q,\Hcal_c\right) & =   \kernel_{p}\left(\Hcal_q,\Hcal_c\right) 
  +\gamma \textsc{Sim}_U\left(\Hcal_q,\Hcal_{c}\right)
\end{align}
where $\gamma$ is a hyperparameter. 
 
\subsection{Neural Parameterization of \our}
Here, we first present the neural architecture of the unwarping function  and then describe the MTPP models used to compute the model based similarity score in Eq.~\eqref{eq:fisher}. As we describe later, we use two MTPP models with different levels of modeling sophistication, \viz, \ourf\ and \ours. In \ourf, the likelihood of a corpus sequence is computed independently of the query sequence using a self attention based MTPP model, \eg, Transformer Hawkes Process~\cite{zuo2020transformer}. As a result, we can employ a locality sensitive hashing based efficient retrieval based \ourf. In \ours, on the other hand, we propose a more expressive and novel cross attention MTPP model, where the likelihood of a corpus sequence is dependent on the query sequence. 
{Thus, our models can effectively tradeoff  between accuracy and efficiency.}

%
%

\xhdr{Neural architecture of $U(\cdot)$}  As discussed in Section~\ref{subsec:components}, the unwarping function $U(\cdot)$
should be unbiased and monotonic.
To this end, we model $U(\cdot)\approx U_{\phi}(\cdot)$ using a nonlinear monotone function which is computed using an unconstrained monotone neural network (UMNN)~\cite{umnn}, \ie,
\begin{align}
 U_{\phi}(t) = \int_0 ^t u_{\phi}(\tau) d\tau + \eta,\label{eq:umnn}
\end{align}
where $\phi$ is the parameter of the underlying neural network $u_{\phi}(\cdot)$, $\eta\in \mathcal{N}(0,\sigma)$ and $u_{\phi}:\RR\to\RR_+$ is a non-negative nonlinear function.
Since the underlying monotonicity can be achieved only by enforcing non-negativity of the integrand $u_{\phi}$, UMNN admits an unconstrained, highly expressive parameterization of monotonic functions. 
Therefore, any complex unwarping function
$U_{\phi}(\cdot)$ can be captured using Eq.~\eqref{eq:umnn}, by integrating a suitable 
neural model augmented with ReLU$(\cdot)$ in the final layer. In other words, if $u_{\phi}$
is a universal approximator for positive function, then $U_{\phi}$ 
can capture any differentiable unwarping function.
We impose an additional regularizer
$\frac{1}{\sigma^2}\int_0 ^T \left\|u_{\phi}(t) -1\right\|^2 dt $ on our training loss  which ensures that
$\left\|\unw(t)-t\right\|$ remains small.

\xhdr{Neural architecture of MTPP model $p_{\theta}(\cdot)$}
We provide two variants of $p_{\theta}(\cdot)$, which leads to two retrieval models, \viz, \ourf\ and \ours. 

\para{\ourf}: Here, we use Transformer Hawkes process~\cite{zuo2020transformer}
which applies a self attention mechanism to model the underlying generative process. In this model, the gradient of corpus sequences
$\vb_{\theta}(\Hcal_c) =\nabla_{\theta} \log p_{\theta}(\Hcal_c)$ are computed independently of the query sequence $\Hcal_q$.
Once we train the retrieval model,  $\vb_{\theta}(\Hcal_c)$ can be pre-computed and bucketized before observing the test query. Such a model, together with the Fisher kernel based cosine similarity scoring model, allows us to apply locality sensitive hashing for efficient retrieval. 
 
\para{\ours}: The above self attention based mechanism models a query agnostic likelihood of the corpus sequences. 
Next, we introduce a \emph{cross attention} based MTPP model 
which explicitly takes into account of the underlying query sequence while modeling the likelihood of the corpus sequence. 
%
Specifically, we measure the latent \eat{cross-attention} {relevance score} between $\Hcal_q$ and $\Hcal_c$ \eat{using the likelihood of a corpus sequence} via a query-induced MTPP model built using the cross attention between the generative process of both the sequences. 
Given a query sequence $\Hcal_q$ and the first $r$ events of  the corpus sequence $\Hcal_c$, we parameterize 
the generative model for $(r+1)$-th event, \ie, $p (e^{(c)}_{r+1} \given  \Hcal({t_r}))$ as $ p_{\theta  \pend}(\cdot)$, where $p_{\theta  \pend} (e^{(c)} _{r+1}) = \intensity _{\theta  \pend} (t^{(c)} _{r+1})\, m^{\pend}_{\theta  \pend} (x^{(c)} _{r+1})$, where $\intensity$ and $m^{\pend}$ are the density and distribution functions for the arrival time and the mark respectively, as described in Eq.~\eqref{eq:qm}. 

\ehdr{Input Layer:} For each event $e_i ^{(q)}$ in the query sequence $\Hcal_q$
and each event $e_j ^{(c)}$ in the first $r$ events in the corpus sequence $\Hcal_c$,  the input layer computes the initial embeddings $\yb^{(q)} _i$ and $\yb^{(c)} _j$ as follows:
\\[-2ex]
\begin{align*}
\yb^{(q)}_i & = \wb_{y, x} x^{(q)}_i + \wb_{y, t}\unw(t^{(q)}_i)  \nn \\
&\quad + \wb_{y, \Delta t} \left(\unw(t^{(q)}_i) - \unw(t^{(q)}_{i-1})\right) + \bb_{y}, \forall i\in[|\Hcal_q|]\\[-1ex]
\yb^{(c)}_j & = \wb_{y, x} x^{(c)}_j + \wb_{y, t}t^{(c)}_j \nn \\
&\quad + \wb_{y, \Delta t} \left(t^{(c)}_j - t^{(c)}_{j-1}\right)   + \bb_{y}, \forall j\in[|\Hcal_c(t_{r})|-1]
\end{align*}
where  
$\wb_{\bullet,\bullet}$ and $\bb_{y}$ are trainable parameters.
%

\ehdr{Attention layer:}
The second layer models the interaction between all the query events and \emph{the past corpus events}, \ie,
$\Hcal_q$ and $\Hcal_c(t_{r})$ using an attention mechanism. 
Specifically, following the existing attention models~\cite{vaswani2017,sasrec,tisasrec} it first adds a trainable position embedding $\pb$ with $\yb$--- the output from the previous layer. 
More specifically, we have the updates:
$\yb^{(q)}_i \leftarrow \yb^{(q)}_i + \pb_i$ and $ \yb^{(c)}_j \leftarrow \yb^{(c)}_j + \pb_j$. 
where, $\pb_\bullet\in \RR^D$. Next, we apply  two linear transformations on the vectors  $[\yb^{(q)}_i]_{i\in [|\Hcal_q|]}$ and one linear transformation on  $[\yb^{(c)}_j]_{j\in[r]}$, \ie,
$ \sb_j = \Wb^S  \yb^{(c)}_j,   \kb_i = \Wb^K  \yb^{(q)}_i,   \vb_i = \Wb^V  \yb^{(q)}_i.$
The state-of-the-art works on attention models~\cite{vaswani2017,zuo2020transformer,sasrec,tisasrec}
often refer $\sb_\bullet$, $\kb_\bullet$ and $\vb_\bullet$ as \querytt,
\keytt\ and \valuett\ vectors respectively. Similarly, we call the trainable weights $\Wb^S, \Wb^K$
and $\Wb^V$ as the \Querytt, \Keytt\ and \Valuett\ matrices, respectively.  
Finally, we use the standard attention recipe~\cite{vaswani2017} to compute the final embedding vector $\hb_j ^{(c,q)}$ for the event $e^{(c)} _j$, induced by query $\Hcal_q$. 
Such a recipe adds the values weighted by the outputs of a softmax function induced by the \querytt\ and \keytt, \ie,
\begin{align}
\hb^{(c,q)} _j = 
 \sum_{i\in [|\Hcal_q|]} \frac{\exp\left( \sb_j ^\top \kb_i /\sqrt{D} \right)}{\sum_{i'\in [|\Hcal_q|]}\exp\left( \sb_j ^\top \kb_{i'} /\sqrt{D} \right)} \vb_i, \label{eq:attn}
\end{align}
%
%
%
\ehdr{Output layer:}
Given the vectors $\hb^{(c,q)} _j$ provided by the attention mechanism~\eqref{eq:attn}, we first
apply a feed-forward neural network on them to compute  $\bhbc_r$ as follows:
\begin{equation*}
\bhbc _{r} = \sum_{j=1}^r \left[\wb_{\overline{h}}\odot\text{ReLU}(\hb^{(c,q)} _j \odot \wb_{h, f} + \bb_{h, o})  + \bb_{\overline{h}}\right],
\end{equation*}
where $\wb_{\bullet,\bullet}$, $\wb_\bullet$ and $\bb_{v}$. Finally, we use these vectors to compute the probability density of the arrival time of the next event $e^{(c)} _{r+1}$, \ie,  $\intensity_{\theta  \pend}(t_{r+1})$ and the mark distribution $m^{\pend} _{\theta  \pend}(x_{r+1})$. In particular, we realize $\intensity_{\theta  \pend}(t_{r+1})$ using a log normal
distribution of inter-arrival times, \ie,
\begin{align}
 \hspace{-2mm}t^{(c)}_{r+1}- t^{(c)}_r \sim \textsc{LogNormal}\left(\mu_e\left(\bhbc _{r} \right),\sigma^2 _e\left(\bhbc _{r} \right) \right)\nn,
\end{align}
where, $\left[\mu_e\left(\bhbc _{r} \right),\sigma_e\left(\bhbc _{r} \right)\right] = \Wb_{t,q}  \bhbc _{r} \hspace{-2mm}+\bb_{t,q}$. Similarly, we model the mark distribution as,
\begin{align}
\hspace{-2mm} m^{\pend} _{\theta  \pend}(x_{r+1}) =  \dfrac{\exp\left(\wb_{x,m}^\top \bhbc +b_{x,m} \right)}{\sum_{x'\in \mathcal{X}}\exp\left(\wb_{x',m}^\top \bhbc +b_{x',m}\right)}, \label{eq:mark-mtpp-x}
\end{align}
where $\Wb_{\bullet,\bullet}$ are the trainable parameters. 
Therefore the set of trainable parameters for the underlying MTPP models is $\theta=\set{\Wb^\bullet, \Wb_{\bullet,\bullet}, \wb_\bullet,\wb_{\bullet,\bullet}, \bb_{\bullet},\bb_{\bullet,\bullet}}$.
%

\subsection{Parameter Estimation}
Given the query sequences $\set{\Hcal_q}$, the corpus sequences $\set{\Hcal_c}$ along with their relevance labels $\set{y(\Hcal_q,\Hcal_c)}$, we seek to find $\theta$ and $\phi$ which ensure that:
\begin{align}
 s_{p_{\theta},\unw_{\phi}}(\Hcal_q,\Hcal_{c_+}) \gg s_{p_{\theta},\unw_{\phi}}(\Hcal_q, \Hcal_{c_-})\, \forall\, c_{\pm} \in \Cr_{q \pm}.
\end{align}
To this aim, we minimize the following pairwise ranking loss~\cite{Joachims2002ranksvm} to estimate the parameters $\theta,\phi$:
\begin{align}
\underset{\theta,\phi} {\text{min}}\sum_{q\in\Qr}\sum_{\substack{c_+\in \Cr_{q\rel},\\ c-\in \Cr_{q\nrel} }}&    \big[s_{p_{\theta},\unw_{\phi}}(\Hcal_q,\Hcal_{c_-})- \nn  s_{p_{\theta},\unw_{\phi}}(\Hcal_q,\Hcal_{c_+}) + \delta \big]_+, 
\end{align}
where, $\delta$ is a tunable margin.


\section{Scalable Retrieval with Hashing} 
\label{sec:hash}
Once we learn the model parameters $\theta$ and $\phi$, we can rank the set of corpus sequences $\Hcal_c$ in the decreasing order of
$s_{p_{\theta},\unw_{\phi}}(\Hcal_{\newq},\Hcal_c)$ for a new query $\Hcal_{\newq}$ and return
top$-K$ sequences. Such an approach requires $|\Ccal|$ comparisons per each test query, which can take a huge amount of time for many real-life applications where $|\Cr|$ is high. 
However, for most practical query sequences, the number of
relevant sequences is a very small fraction of the entire corpus of sequences. Therefore, the number of comparisons between query-corpus sequence pairs can be reduced without significantly impacting the retrieval quality by selecting a small number of candidate corpus sequences that are more likely to be relevant to a query sequence.

\subsection{Trainable Hashing for Retrieval}

\xhdr{Computation of a trainable hash code}  We first apply a trainable
nonlinear transformation $\Lambda_{\psi}$ with parameter $\psi$ on the gradients $\zzc = \vb_{p_{\theta}}(\Hcal_c)$ and then learn the binary hash vectors $\hash^c = \sgn\bracex{\Lambda_{\psi}\bracex{\zzc}}$  by solving the following optimization, where we use $\tanh\left(\Lambda_{\psi}(\cdot)\right)$
as a smooth approximation of $\sgn\left(\Lambda_{\psi}(\cdot)\right)$.
\begin{align}\label{eq:hash}
 & \min_{\psi} \frac{\eta_1}{|\Cr|}\sum_{c\in\Cr}   \left|\bm{1}^\top \tanh\bracex{\Lambda_{\psi}\bracex{\zzc}}\right| \nn\\[-2ex]
&\qquad\qquad\qquad  + \frac{\eta_2}{|\Cr|}\sum_{c\in\Cr}   \left\| \left|\tanh\bracex{\Lambda_{\psi}\bracex{\zzc}}\right|-\bm{1}\right\|_1  \\[-2ex]  
 &\qquad + \frac{2\eta_3}{{D \choose 2}}  \cdot \bigg| \sum_{\substack{c\in \Ccal \\ i\neq j\in [D]}}  \tanh\bracex{\Lambda_{\psi}\bracex{\zzc}[i]} \cdot  \tanh\bracex{\Lambda_{\psi} \bracex{\zzc}[j]}\bigg| \nn
 \end{align}
\setlength{\textfloatsep}{2pt}
\begin{algorithm}[t]
\small
        \caption{Efficient retrieval with hashing}
        \begin{algorithmic}[1]
          \REQUIRE  Trained corpus embeddings
          $\set{\vb^c = \vb_{p_{\theta}} (\Hcal_c)} $ using \ourf; new query sequences $\set{\Hcal_{\newq}}$, $K$:
          \# of corpus sequences to return; trained models for \ourf\ and \ours.
          \STATE \textbf{Output:} $\set{L_\newq}$: top-$K$ relevant sequences from $\set{\Hcal_c}$.
          \STATE $\psi\leftarrow\textsc{TrainHashNet}\left(\Lambda_{\psi},[\vb^c]_{c\in\Cr} \right)$
          \STATE \textsc{InitHashBuckets}$(\cdot )$ 
          \FOR{$c\in \Cr$} 
          \STATE $\hash^c\leftarrow \textsc{ComputeHashCode}\left(\zzc;\Lambda_{\psi}\right)$
          \STATE $\Bcal\leftarrow\textsc{AssignBucket}(\hash^c)$
          \ENDFOR
          \FOR{each new query $\Hcal_{\newq}$} 
          \STATE $\vb^{q'}\leftarrow  \ourf(\Hcal_{\newq})$ 
          \STATE $\hash^{q'}\leftarrow \textsc{ComputeHashCode} (\vb^{q'};  {\Lambda_{\psi}} )$ 
          \STATE  $\Bcal\leftarrow\textsc{AssignBucket}(\hash^{\newq})$ 
            \FOR{$c\in \Bcal$} 
          \STATE $\vb^{\newq} _{\text{cross}}, \vb^{c} _{\text{cross}} \leftarrow \ours(\Hcal_{\newq},\Hcal_c)$
          \STATE $s_{p_{\theta},U_{\phi}} (\Hcal_{q'},\Hcal_c) \leftarrow \textsc{Score}(\vb^{\newq} _{\text{cross}}, \vb^{c} _{\text{cross}}, \Hcal_q,\Hcal_c)$
          \ENDFOR

          \STATE $L_{\newq}\leftarrow \textsc{Rank}(\set{s_{p_{\theta},U_{\phi}} (\Hcal_{q'},\Hcal_c)},K)$
        \ENDFOR
        \STATE \textbf{Return}  $\set{L_{\newq}}$
        \end{algorithmic}      \label{alg:key}
      \end{algorithm}
Here $\sum_{i=1}^3 \eta_i=1$. Moreover, different terms in Eq.~\eqref{eq:hash} allow the hash codes $\hash^c$ to have a set of four desired properties: (i) the first term ensures that the numbers of $+1$ and $-1$ are evenly distributed in the hash vectors
 $\hash^c=  \tanh\bracex{\Lambda_{\psi}\bracex{\zzc}}$; (ii) the second term encourages the entries of $\hash^c$ become as close to $\pm 1$ as possible so that $\tanh(\cdot)$ gives an accurate approximation of $\sgn(\cdot)$ ; (iii) the third term ensures that  the entries of the hash codes
 $\hash^c$ contain independent information and therefore they have no redundant entries.  Trainable hashing has been used in other domains of information retrieval including graph hashing~\cite{liu2014discrete,qin2020ghashing,roy2020adversarial}, document retrieval~\cite{zhang2010self, salakhutdinov2009semantic, dadaneh20a}. However, to the best of our knowledge, such an approach has never been proposed for continuous time sequence retrieval.
 
\xhdr{Outline of our retrieval method}
We summarize our retrieval procedure in Algorithm~\ref{alg:key}.  
We are given gradient vectors $\vb^c = \vb_{p_{\theta}}(\Hcal_c)$ obtained by training \ourf.  Next, we  train an additional neural network $\Lambda_{\psi}$ parameterized by $\psi$ (\textsc{TrainHashNet}$()$, line 2), which is used to learn a binary hash vector $\hash^c$ for each sequence $\Hcal_c$.
Then these hash codes are used to arrange corpus sequences in different hash buckets (for-loop in lines 4--7) using the algorithm proposed by~\citet{GionisIM1999hash}, so that two sequences $\Hcal_c, \Hcal_{c'}$ lying in the same
hash bucket have very high value of cosine similarity $\cos(\vb^c,\vb^{c'})$ (bucketization details are in Appendix B).
%
%
Finally, once a new query $\Hcal_\newq$ comes, we first compute  $\vb^{\newq}$
using the trained \ourf\ model and then compute the binary hash codes 
$\hash^{\newq}$ using the trained hash network $\Lambda_\psi$ (lines 9--10). Next, we assign an appropriate bucket $\Bcal$ to it (line 11) and finally compare it with only the corpus sequences in the same bucket, \ie, $\Hcal_c\in\Bcal$ (lines 12--15) using our model. 

Recall that we must use \ourf\ to compute gradient vectors for subsequent hash code generation (in lines 9--10). However, at the last stage for final score computation and ranking,  we can use any variant of \our (in line 13), preferably \ours, since the corpus sequences have already been indexed by our LSH method.

\section{Experiments}

In this section, we provide a comprehensive evaluation of \our and our hashing method.  
Appendix D~\cite{tppsel} contains additional experiments. Our implementation and datasets are available at \texttt{https://github.com/data-iitd/neuroseqret/}.

\newcommand{\secbest}{\underline}
\newcommand{\thbest}{\fbox}

\begin{table*}[h]
\setlength{\tabcolsep}{3pt}
\centering
\resizebox{0.84\textwidth}{!}{\tabcolsep 6pt
\hspace{-8mm}
\begin{tabular}{l|ccccc|ccccc}
\toprule
 & \multicolumn{5}{c|}{\textbf{Mean Average Precision (MAP) in \%}} & \multicolumn{5}{c}{\textbf{NDCG@10 in \%}} \\ \hline 
 & Audio & Celebrity & Electricity & Health & Sports & Audio & Celebrity & Electricity & Health & Sports \\ \hline \hline
MASS~\cite{mass} & 51.1$\pm$0.0 & 58.2$\pm$0.0 & 19.3$\pm$0.0 & 26.4$\pm$0.0 & 54.7$\pm$0.0 & 20.7$\pm$0.0 & 38.7$\pm$0.0 & 9.1$\pm$0.0 & 13.6$\pm$0.0 & 22.3$\pm$0.0 \\
UDTW~\cite{ucrdtw} & 50.7$\pm$0.0 & 58.7$\pm$0.0 & 20.3$\pm$0.0 & 28.1$\pm$0.0 & 54.5$\pm$0.0 & 21.3$\pm$0.0 & 39.6$\pm$0.0 & 9.7$\pm$0.0 & 14.7$\pm$0.0 & 22.9$\pm$0.0 \\
Sharp~\cite{blondel2021differentiable} & 52.4$\pm$0.2 & 59.8$\pm$0.5 & 22.8$\pm$0.2 & 28.6$\pm$0.2 & \thbest{56.8$\pm$0.3} & 21.9$\pm$0.2 & 40.6$\pm$0.5 & 11.7$\pm$0.1 & 16.8$\pm$0.1 & 23.7$\pm$0.2 \\
RMTPP~\cite{du2016recurrent} & 48.9$\pm$2.3 & 57.6$\pm$1.8 & 18.7$\pm$0.8 & 24.8$\pm$1.2 & 50.3$\pm$2.5 & 20.1$\pm$1.9 & 39.4$\pm$2.1 & 8.3$\pm$0.8 & 12.3$\pm$0.5 & 19.1$\pm$1.8 \\
\tpprank-RMTPP & 52.6$\pm$2.0 & 60.3$\pm$1.7 & 23.4$\pm$0.7 & 29.3$\pm$0.6 & 55.8$\pm$2.1 & 22.4$\pm$1.3 & 41.2$\pm$1.3 & 11.4$\pm$0.4 & 15.5$\pm$0.5 & 23.9$\pm$1.4 \\
SAHP~\cite{zhang2019self} & 49.4$\pm$3.2& 57.2$\pm$2.9 & 19.0$\pm$1.8 & 26.0$\pm$2.1 & 53.9$\pm$3.6 & 20.4$\pm$2.3 & 39.0$\pm$3.1 & 8.7$\pm$1.2 & 13.2$\pm$1.4 & 22.6$\pm$2.5 \\
\tpprank-SAHP & 52.9$\pm$1.8& 61.8$\pm$2.3 & 26.5$\pm$1.2 & 31.6$\pm$1.1 & 55.1$\pm$2.3 & 23.3$\pm$1.4 & 42.1$\pm$1.7 & 13.3$\pm$0.7 & 17.5$\pm$0.9 & 25.4$\pm$1.8 \\
THP~\cite{zuo2020transformer} & 51.8$\pm$2.3& 60.3$\pm$1.9 & 21.3$\pm$0.9 & 27.9$\pm$0.9 & 54.2$\pm$2.1 & 22.1$\pm$1.1 & 40.3$\pm$1.2 & 10.4$\pm$0.6 & 14.4$\pm$0.3 & 22.9$\pm$1.1 \\
\tpprank-THP & \thbest{54.3$\pm$1.7}& \thbest{63.1$\pm$2.1} & \thbest{29.4$\pm$0.9} & \thbest{33.6$\pm$1.3} &  56.3$\pm$1.9  & \thbest{25.4$\pm$0.9} & \thbest{44.2$\pm$1.0} & \thbest{15.3$\pm$0.4} & \thbest{19.7$\pm$0.4} & \thbest{26.5$\pm$0.9} \\ \hline
\ourf & \secbest{55.8$\pm$1.8} & \secbest{64.4$\pm$1.9} & \secbest{30.7$\pm$0.7} & \secbest{35.9$\pm$0.9} & \secbest{57.6$\pm$1.9} & \secbest{25.9$\pm$1.1} & \secbest{45.8$\pm$1.0} & \secbest{16.5$\pm$0.5} & \secbest{20.4$\pm$0.4} & \secbest{27.8$\pm$1.1} \\
\ours & \textbf{56.2$\pm$2.1}\textsuperscript{$\dagger$} & \textbf{65.1$\pm$1.9}\textsuperscript{$\dagger$} & \textbf{32.4$\pm$0.8}\textsuperscript{$\dagger$} & \textbf{37.4$\pm$0.9}\textsuperscript{$\dagger$} & \textbf{58.7$\pm$2.1} & \textbf{28.3$\pm$1.1}\textsuperscript{$\dagger$} & \textbf{46.9$\pm$1.2}\textsuperscript{$\dagger$} & \textbf{18.1$\pm$0.7}\textsuperscript{$\dagger$} & \textbf{22.0$\pm$0.4}\textsuperscript{$\dagger$} & \textbf{27.9$\pm$1.2} \\
\bottomrule
\end{tabular}
}\\[-1ex]
\caption{Retrieval accuracy in terms of mean average precision (MAP) and NDCG@10 (both in \%) of  all the methods across five datasets on the test set. Numbers with bold font (underline) indicate best (second best) performer. \textbf{Boxed} numbers indicate best performing state-of-the-art baseline. Results marked \textsuperscript{$\dagger$} are statistically significant (two-sided Fisher's test with $p \le 0.05$) over the best performing state-of-the-art baseline (\tpprank-THP or Sharp).
The standard deviation for MASS and UDTW are zero, since they are deterministic retrieval algorithms.\\[-2ex]}
\label{tab:main}
\end{table*}
\newcommand{\seq}{\texttt{seq}}

\subsection{Experimental Setup} \label{sec:exp-setup}
\xhdr{Datasets} 
We evaluate the retrieval performance of \our and other methods across large-scale real-world datasets with up to 60 million events. The statistics of all datasets are given in Appendix~\cite{tppsel}. Across all datasets, $|\Hcal_q| = 5K$ and $|\Hcal_c| = 200K$.  

\begin{asparaenum}[(1)]
\item \textbf{Audio:} The dataset contains audio files for spoken commands to a smart-light system and the demographics(age, nationality) of the speaker.  Here, a query corpus sequence pair is relevant if they are from an audio file with a common speaker.

\item \textbf{Sports:} The dataset contains actions (\eg run, pass, shoot) taken while playing different sports. We consider the time of action and action class as time and mark of sequence respectively. Here, a query corpus sequence pair is relevant if they are from a common sport.

\item \textbf{Celebrity:} In this dataset, we consider the series of frames extracted from youtube videos of multiple celebrities as event sequences where event-time denotes the video-time and the mark is decided upon the coordinates of the frame where the celebrity is located.
Here, a query corpus sequence pair is relevant if they are from a video file having a common celebrity.

\item \textbf{Electricity:}  This dataset contains the power-consumption records of different devices across smart-homes in the UK. We consider the records for each device as a sequence with event mark as the \textit{normalized} change in the power consumed by the device and the time of recording as event time. 
Here, a query corpus sequence pair is relevant if they are from a  similar appliance.

\item \textbf{Health:} The dataset contains ECG records for patients suffering from heart-related problems. Since the length of the ECG record for a single patient can be up to 10 million, we generate smaller individual sequences of length 10,000 and consider each such sequence as an independent sequence. The marks and times of events in a sequence are determined using a similar procedure as in Electricity. 
A query corpus sequence pair is relevant if they are from a common patient.
\end{asparaenum}

For Health, Celebrity and Electricity, we lack the true ground-truth labeling of relevance between sequences. Therefore, we adopt a heuristic in which, given a dataset $\Dcal$, from each sequence $\seq_q\in\Dcal$ with $q\in[|\Dcal|]$, we first sample a set of sub-sequences $\mathcal{U}_q=\set{\Hcal\subset \texttt{seq}_q}$ with $|\Ucal_q|\sim \text{Unif}\,[200,300]$. For each such collection $\Ucal_q$, we draw exactly one query $\Hcal_q$ uniformly at random from $\Ucal_q$, \ie, $\Hcal_q\sim\Ucal_q$. Then, we define $\Cr=\cup_{q\in[|\Dcal|]}\Ucal_q\cp \Hcal_q$, $\Cr_{q\rel}=\Ucal_q\cp \Hcal_q$ and $\Cr_{q\nrel}=\cup_{c\neq q}\big(\Ucal_c\cp \Hcal_c\big)$.


\xhdr{Baselines} We consider three continuous time-series retrieval models: (i) MASS~\cite{mass}, (ii) UDTW~\cite{ucrdtw} and (iii) Sharp~\cite{blondel2021differentiable}; and, three MTPP models (iv) RMTPP~\cite{du2016recurrent}, (v) SAHP~\cite{zhang2019self}, and (vi) THP~\cite{zuo2020transformer}.
For sequence retrieval with MTPP models, we first train them across all the sequences using maximum likelihood estimation.
Then, given a test query $\Hcal_{\newq}$, this MTPP method ranks the corpus sequences $\set{\Hcal_c}$
in decreasing order of their cosine similarity  $\text{CosSim}(\texttt{emb}^{(\newq)},\texttt{emb}^{(c)})$, where $\texttt{emb}^{(\bullet)}$ is the corresponding sequence embedding provided by the underlying MTPP model.  
In addition, we build supervised ranking models over these approaches, \emph{viz.,} \tpprank-RMTPP, \tpprank-SAHP and \tpprank-THP corresponding to RMTPP, SAHP, and THP. Specifically, \tpprank-MTPP formulates a ranking loss on the query-corpus pairs based on the cosine similarity scores along with the likelihood function to get the final training objective. Therefore, the vanilla MTPP models are used as unsupervised models and the corresponding \tpprank-MTPP models work as supervised models.  
Appendix C~\cite{tppsel} contains the implementation details.

\xhdr{Evaluation protocol} We partition the set of queries $\Qr$ into 50\% training, 10\% validation and rest as test sets. First, we train a retrieval model using the set of training queries. Then, for each test query $\newq$, we use the trained   model to obtain a top-$K$ ranked list from the corpus sequences. Next,  we compute the average precision (AP) and discounted cumulative gain (DCG) of each top-$K$ list, based on the   ground truth. Finally, we compute the mean average precision (MAP) and NDCG@$K$, by averaging AP and DCG values across all test queries. We set $K=10$. 
 
\subsection{Results on Retrieval Accuracy}
\xhdr{Comparison with baselines}
First, we compare our model against the baselines in terms of MAP~and~NDCG.  
Table~\ref{tab:main} summarizes the results, which shows that (i) both \ours\ and \ourf\ outperform all the baselines by a substantial margin;
(ii) \ours outperforms \ourf, since the former has a higher expressive power; 
(iii) the variants of baseline MTPP models trained for sequence retrieval, \ie, \tpprank-RMTPP, \tpprank-SAHP, and \tpprank-THP outperform the vanilla MTPP models; 
(iv) the performances of vanilla MTPPs and the time series retrieval models (MASS, UDTW and Sharp) are comparable.
 
\begin{table}[t]
	\small
	\centering
	\resizebox{0.49\textwidth}{!}{
	\begin{tabular}{l|ccc}
	\toprule
	\textbf{Variant} & \textbf{Audio} & \textbf{Celebrity} & \textbf{Health}\\ \hline
	(i) $s_{p_{\theta},U_{\phi}} (\Hcal_q,\Hcal_c) = -\Delta_x (\Hcal_q,\Hcal_c) - \Delta_t (U_{\phi}(\Hcal_q),\Hcal_c)$ & 36.1$\pm$0.0 & 43.7$\pm$0.0 & 18.9$\pm$0.0\\
	(ii) $s_{p_{\theta},U_{\phi}} (\Hcal_q,\Hcal_c)= \kernel_{p_{\theta}}(\Hcal_q,\Hcal_c)$ & 53.9$\pm$1.9 & 62.5$\pm$1.3 & 33.6$\pm$0.7\\
	(iii) $s_{p_{\theta},U_{\phi}} (\Hcal_q,\Hcal_c)= \kernel_{p_{\theta}}(\Hcal_q,\Hcal_c) -\gamma \Delta_x (\Hcal_q,\Hcal_c)$ & 54.6$\pm$1.9 & 63.1$\pm$1.4 & 33.7$\pm$0.7\\
(iv)	$s_{p_{\theta},U_{\phi}} (\Hcal_q,\Hcal_c)= \kernel_{p_{\theta}}(\Hcal_q,\Hcal_c) -\gamma \Delta_t (U_{\phi}(\Hcal_q),\Hcal_c)$ & 55.7$\pm$2.0 & 63.7$\pm$1.8 & 35.9$\pm$0.8\\
	(v) \ours Without $U_{\phi}(\cdot)$ & 55.2$\pm$2.2 & 62.9$\pm$2.0 & 34.3$\pm$0.9\\
	(vi) \ours & 56.2$\pm$2.1 & 65.1$\pm$1.9 & 37.4$\pm$0.9\\
	\bottomrule
	\end{tabular}
	}\\[-1.5ex]
	\caption{Ablation study.}
	\vspace{-2mm}
	\label{tab:model_ablation}
\end{table}
\begin{figure}[h]
 \centering
\subfloat[$\Hcal_q, \Hcal_c$]
{\includegraphics[height=1.7cm]{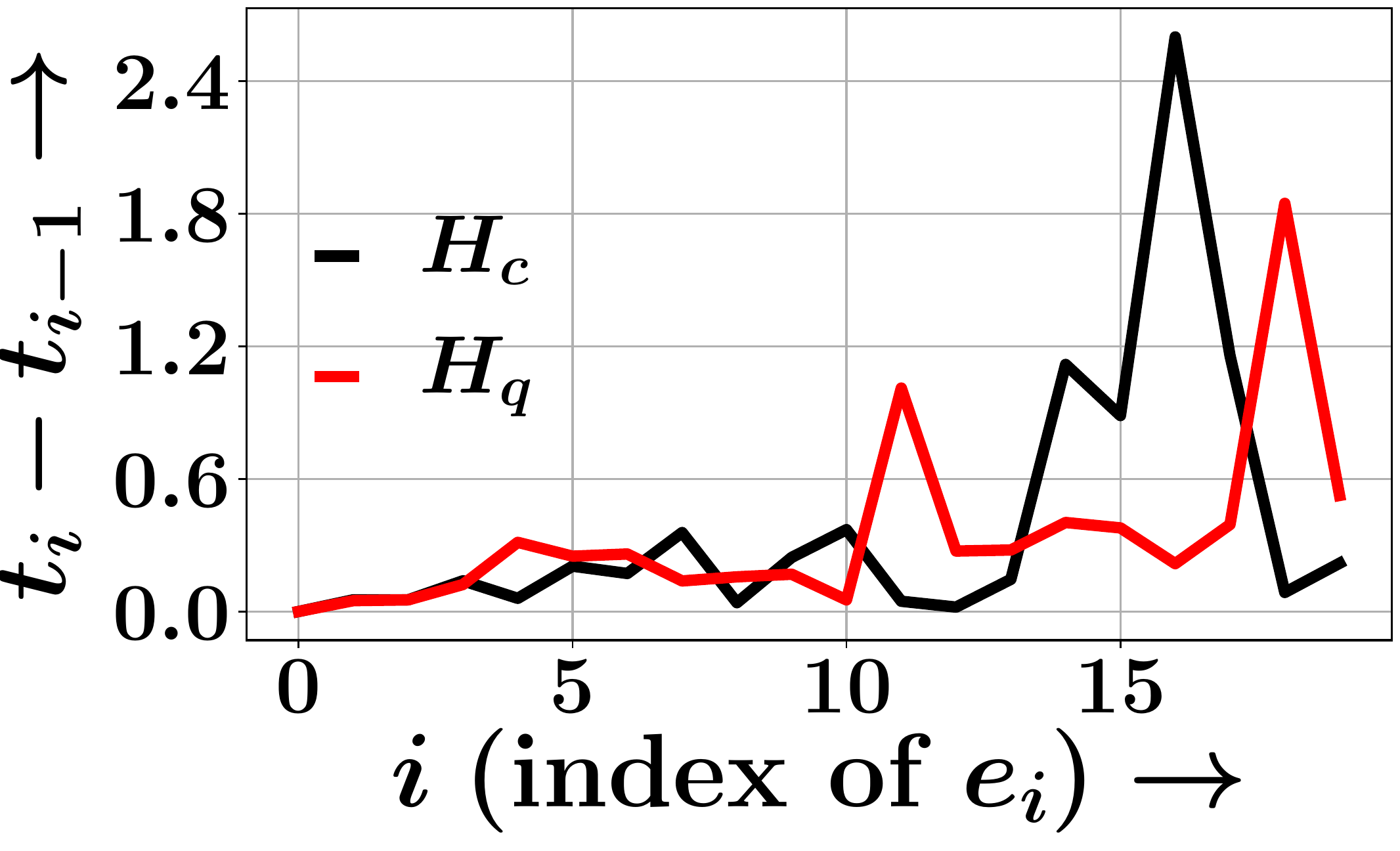}}
\hspace{2mm}
\subfloat[$U_{\phi}(\Hcal_q), \Hcal_c$]
{\includegraphics[height=1.7cm]{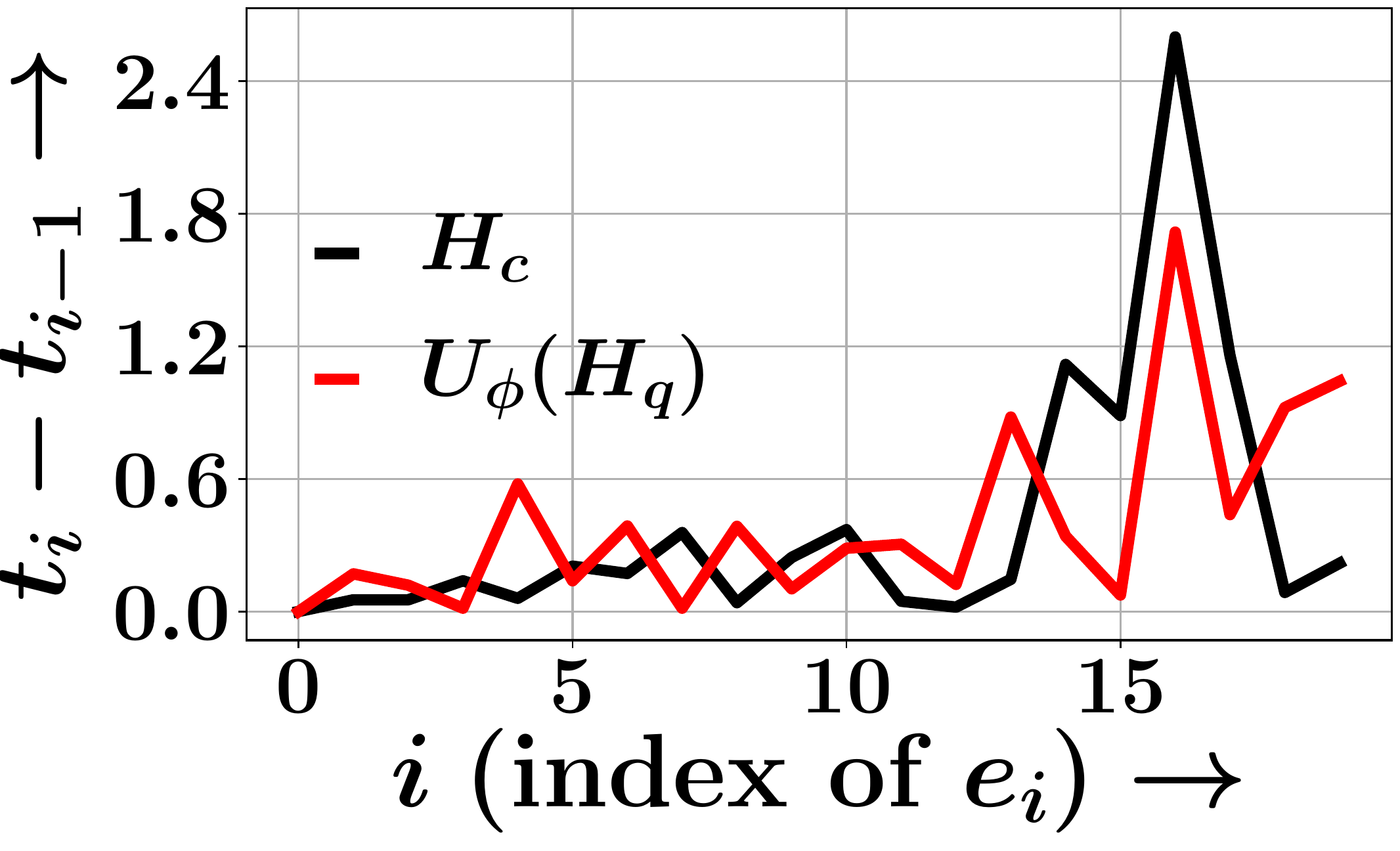}}\\[-1ex]
\caption{Effect of unwarping on a \emph{relevant} query-corpus pair in Audio. $U_{\phi}(\cdot)$ learns to transform $\Hcal_q$ in order to capture a high value of its latent similarity with $\Hcal_c$.}
\label{fig:UU}
\end{figure}

\xhdr{Ablation study}
Next, we compare the retrieval performance across four model variants: 
\begin{inparaenum}[(i)] 
\item our model with only model-independent score \ie, $s_{p_{\theta},U_{\phi}} (\Hcal_q,\Hcal_c) = -\Delta_x (\Hcal_q,\Hcal_c) - \Delta_t (U_{\phi}(\Hcal_q),\Hcal_c)$;
 \label{var:withoutmodel}
\item our model with only model-dependent score, \ie, $s_{p_{\theta},U_{\phi}}(\Hcal_q,\Hcal_c) = \kernel_{p_{\theta}}(\Hcal_q,\Hcal_c)$;
 \label{var:withmodel}
\item our model without any model independent time similarity \ie, $s_{p_{\theta},U_{\phi}}(\Hcal_q,\Hcal_c) = \kernel_{p_{\theta}}(\Hcal_q,\Hcal_c) -\gamma \Delta_x (\Hcal_q,\Hcal_c)$;
 \label{var:withouttime}
\item our model without any model independent mark similarity \ie, $s_{p_{\theta},U_{\phi}}(\Hcal_q,\Hcal_c) = \kernel_{p_{\theta}}(\Hcal_q,\Hcal_c) -\gamma \Delta_t (U_{\phi}(\Hcal_q),\Hcal_c)$;
 \label{var:withoutmark}
\item  our model without unwarping function $U_{\phi}(\cdot)$;
 \label{var:withoutU}
 and
\item the complete design of our model.
\label{var:our}
\end{inparaenum}
In all cases, we used \ours.  

Table~\ref{tab:model_ablation} shows that 
the complete design of our model~(variant \eqref{var:our})  achieves the best performance.  
We further note that removing $\kernel_{p_\theta}$ from the score (variant \eqref{var:withoutmodel}) leads to significantly poor performance.
Interestingly, our model without any mark based similarity~(variant \eqref{var:withoutmark}) leads to better performance than the model without time similarity  (variant~\eqref{var:withouttime})--- this could be attributed to the larger variance in query-corpus time distribution than the distribution of marks. 
Finally, we observe that the performance deteriorates if we do not use an unwarping function $U_{\phi}(\cdot)$ (variant~\eqref{var:withoutU}). 
Figure~\ref{fig:UU} illustrates the effect of $U_{\phi}(\cdot)$. It shows that
 $U_{\phi}(\cdot)$ is able to learn suitable transformation of the query sequence, which encapsulates the high value of latent similarity with the corpus sequence.

\begin{figure}[t]
\centering
{\includegraphics[height=0.7cm]{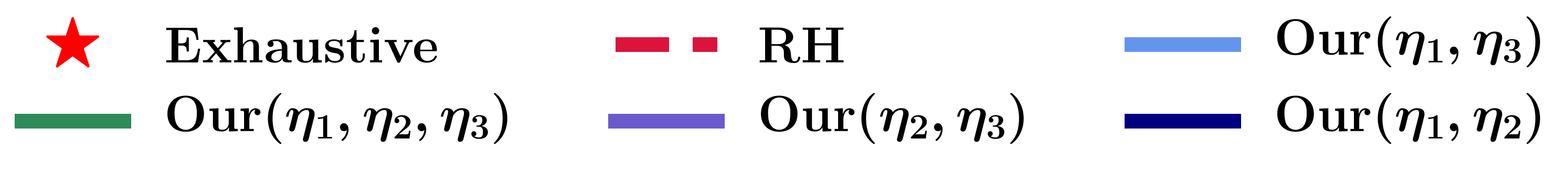}}\\[-2.5ex]
\subfloat
{\includegraphics[height=1.8cm]{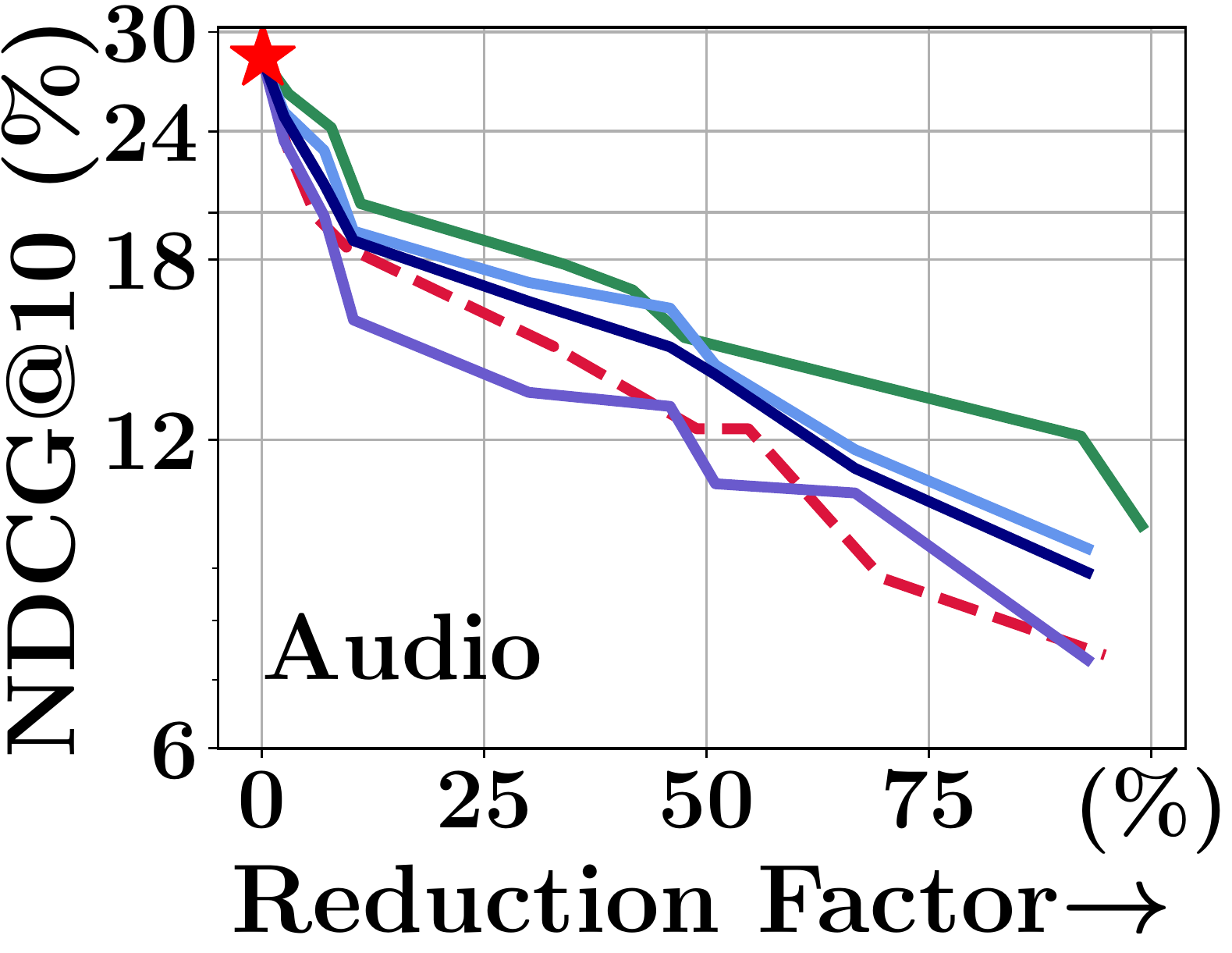}}
\hfill
\subfloat
{\includegraphics[height=1.8cm]{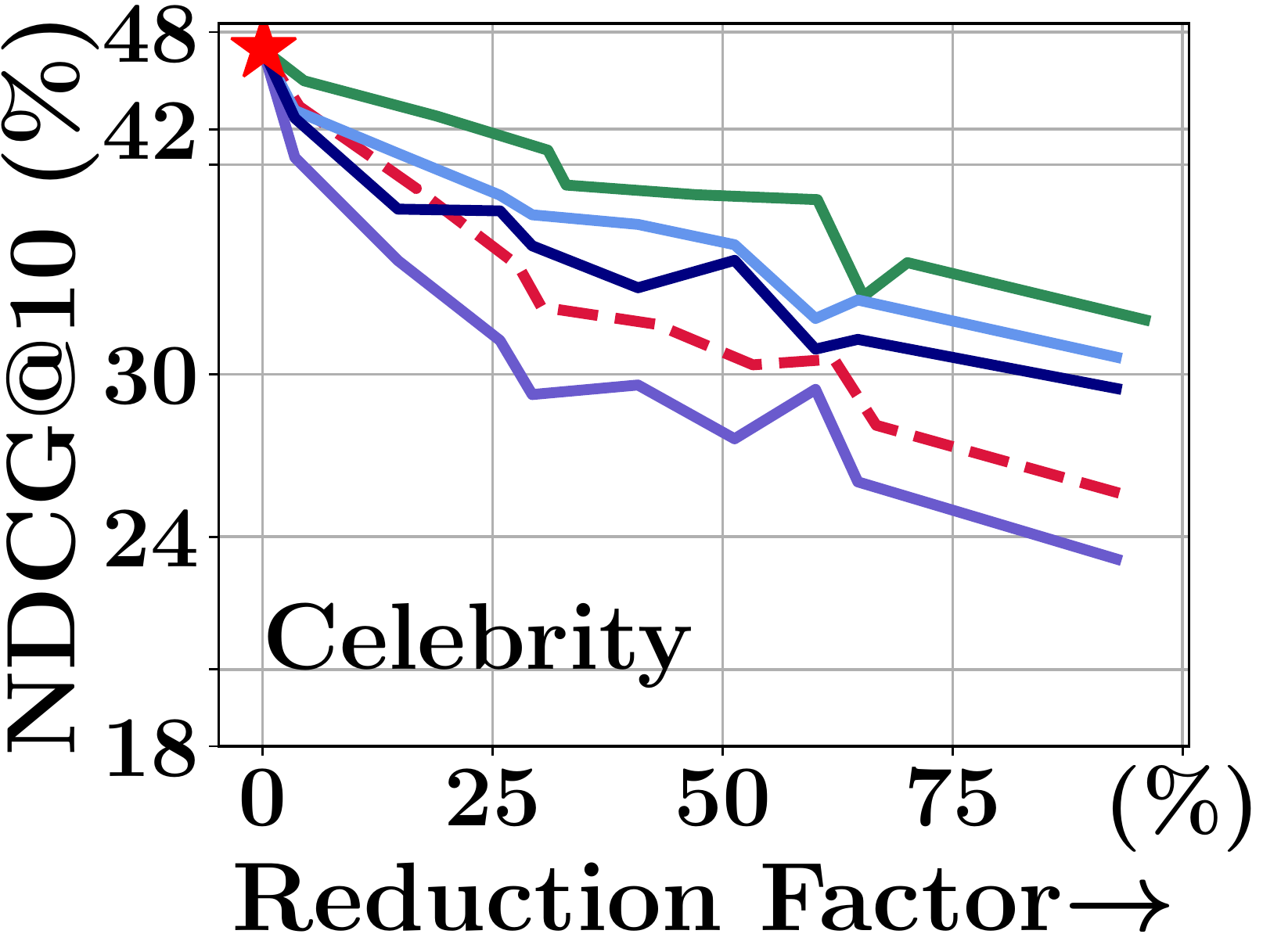}}
\hfill
\subfloat
{\includegraphics[height=1.8cm]{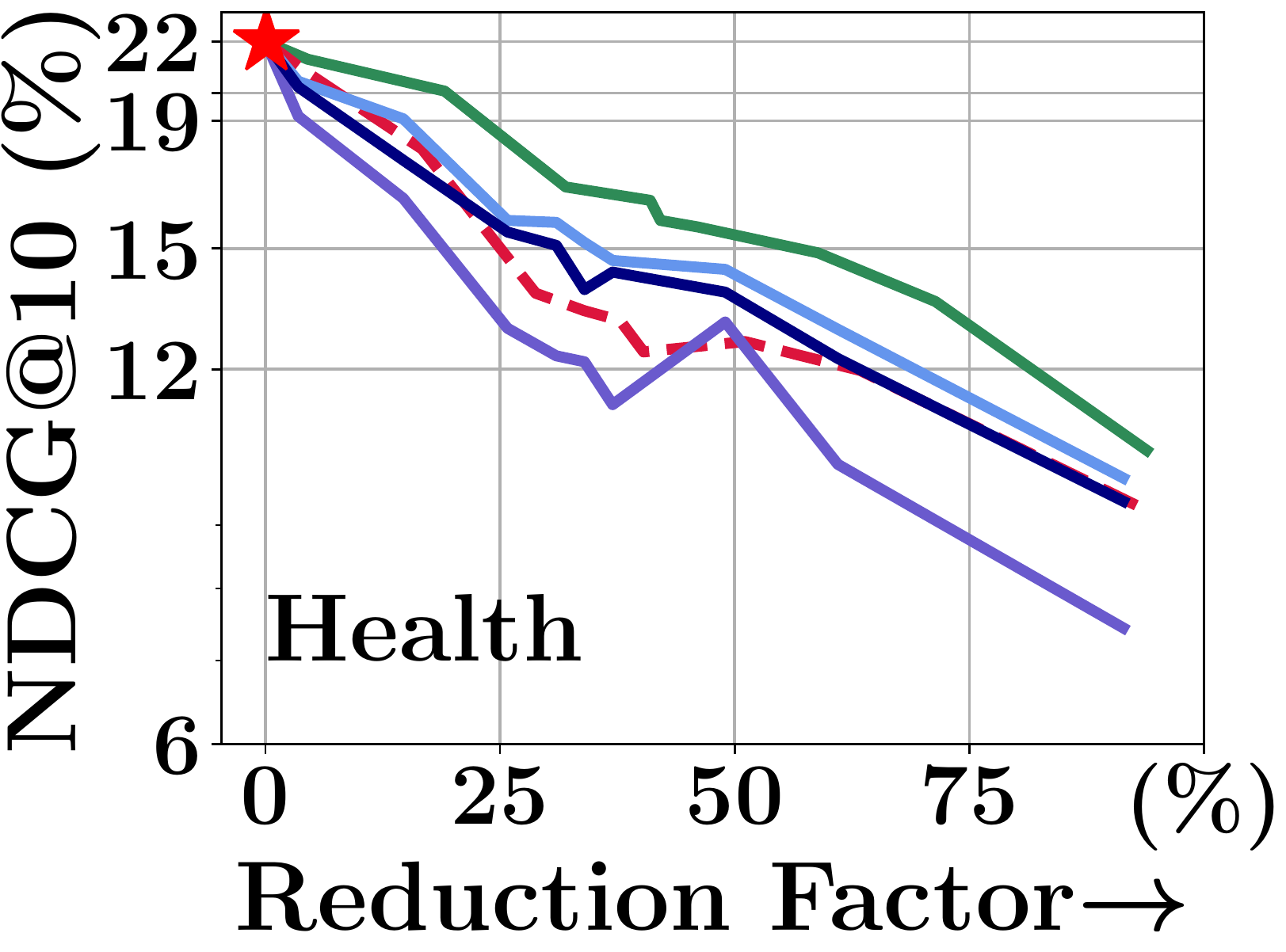}}
\caption{Tradeoff between NDCG@10 vs. Reduction factor, \ie, \% reduction in number of comparisons between query-corpus pairs w.r.t. the exhaustive comparisons for different hashing methods. The point marked as $ \color{red} {\star}$ indicates the case with exhaustive comparisons on the set of corpus sequences. }
\label{fig:hash}
\end{figure}

\subsection{Results on Retrieval Efficiency}
We compare our efficient sequence retrieval method  given in Algorithm~\ref{alg:key} against random hyperplane (RH) method (Appendix B in \cite{tppsel})
and three variants of our proposed training problem in Eq.~\eqref{eq:hash}. 
\begin{inparaenum}[(i)] 
\item Our$(\eta_2, \eta_3)$ which sets $\eta_1=0$ and thus does not enforce even distribution of $\pm 1 $ in $\hash^c$;
\item Our$(\eta_1, \eta_3)$ which sets $\eta_2=0$ and thus $\tanh$ does not accurately approximate $\sgn$;
\item Our$(\eta_1, \eta_2)$ which sets $\eta_3=0$ and thus does not enforce $\hash^c$ to be compact and free of redundancy. 
\end{inparaenum}
Our$(\eta_1,\eta_2,\eta_3)$ is the complete design which includes all trainable components. Figure~\ref{fig:hash} summarizes the results.

\xhdr{Comparison with random hyperplane}   Figure~\ref{fig:hash}  shows that our method (Our$(\eta_1,\eta_2,\eta_3)$) demonstrates better Pareto efficiency than RH. 
This is because RH generates hash code in a data oblivious manner whereas our method learns the hash code on top of the trained embeddings.

\xhdr{Ablation study on different components of Eq.~\eqref{eq:hash}}
Figure~\ref{fig:hash} summarizes the results, which shows that (i) the first three variants are outperformed by Our$(\eta_1,\eta_2,\eta_3)$; (ii) 
the first term having $\eta_1 \neq 0$, which enforces an even distribution of $\pm 1$, is the most crucial component for the loss function--- as the removal of this term
causes significant deterioration of the performance.

\section{Conclusions}
\label{sec:conclusions}
In this paper, we proposed a novel supervised continuous time event sequence retrieval system called \our\ using neural MTPP models.  To achieve efficient retrieval over very large corpus of sequences, we also propose a trainable hash-coding of corpus sequences which can be used to narrow down the number of sequences to be considered for similarity score computation. Our experiments with real world datasets from a diverse range of domains show that both our retrieval model and hashing methods are more effective than several baselines. 

\clearpage

\bibliography{refs} 

\begin{thebibliography}{58}
\providecommand{\natexlab}[1]{#1}

\bibitem[{Abanda et~al.(2019)Abanda, Mori, and Lozano}]{dtw_paparrizos}
Abanda, A.; Mori, U.; and Lozano, J.~A. 2019.
\newblock A review on distance based time series classification.
\newblock In \emph{DMKD}.

\bibitem[{Alaee et~al.(2020)Alaee, Kamgar, and Keogh}]{alaee2020matrix}
Alaee, S.; Kamgar, K.; and Keogh, E. 2020.
\newblock Matrix Profile XXII: Exact Discovery of Time Series Motifs under DTW.
\newblock In \emph{ICDM}.

\bibitem[{Blondel et~al.(2021)Blondel, Mensch, and
  Vert}]{blondel2021differentiable}
Blondel, M.; Mensch, A.; and Vert, J.-P. 2021.
\newblock Differentiable Divergences Between Time Series.
\newblock In \emph{AISTATS}.

\bibitem[{Cai et~al.(2019)Cai, Xu, Yi, Huang, and Rajasekaran}]{cai2019dtwnet}
Cai, X.; Xu, T.; Yi, J.; Huang, J.; and Rajasekaran, S. 2019.
\newblock DTWNet: a dynamic time warping network.
\newblock In \emph{NeurIPS}.

\bibitem[{Charikar(2002)}]{charikar2002similarity}
Charikar, M.~S. 2002.
\newblock Similarity estimation techniques from rounding algorithms.
\newblock In \emph{STOC}.

\bibitem[{Cuturi and Blondel(2017)}]{cuturi2017soft}
Cuturi, M.; and Blondel, M. 2017.
\newblock Soft-dtw: a differentiable loss function for time-series.
\newblock In \emph{ICML}.

\bibitem[{Daley and Vere-Jones(2007)}]{daley2007introduction}
Daley, D.~J.; and Vere-Jones, D. 2007.
\newblock \emph{An Introduction to the Theory of Point Processes: Volume II:
  General Theory and Structure}.
\newblock Springer Science \& Business Media.

\bibitem[{De et~al.(2018)De, Bhattacharya, and Ganguly}]{de2018demarcating}
De, A.; Bhattacharya, S.; and Ganguly, N. 2018.
\newblock Demarcating endogenous and exogenous opinion diffusion process on
  social networks.
\newblock In \emph{WWW}.

\bibitem[{De et~al.(2016)De, Valera, Ganguly, Bhattacharya, and
  Gomez-Rodriguez}]{de2016learning}
De, A.; Valera, I.; Ganguly, N.; Bhattacharya, S.; and Gomez-Rodriguez, M.
  2016.
\newblock Learning and Forecasting Opinion Dynamics in Social Networks.
\newblock In \emph{NeurIPS}.

\bibitem[{Du et~al.(2016)Du, Dai, Trivedi, Upadhyay, Gomez-Rodriguez, and
  Song}]{du2016recurrent}
Du, N.; Dai, H.; Trivedi, R.; Upadhyay, U.; Gomez-Rodriguez, M.; and Song, L.
  2016.
\newblock Recurrent marked temporal point processes: Embedding event history to
  vector.
\newblock In \emph{KDD}.

\bibitem[{Du et~al.(2015)Du, Farajtabar, Ahmed, Smola, and
  Song}]{du2015dirichlet}
Du, N.; Farajtabar, M.; Ahmed, A.; Smola, A.~J.; and Song, L. 2015.
\newblock Dirichlet-hawkes processes with applications to clustering
  continuous-time document streams.
\newblock In \emph{KDD}.

\bibitem[{Farajtabar et~al.(2017)Farajtabar, Yang, Ye, Xu, Trivedi, Khalil, Li,
  Song, and Zha}]{farajtabar2017fake}
Farajtabar, M.; Yang, J.; Ye, X.; Xu, H.; Trivedi, R.; Khalil, E.; Li, S.;
  Song, L.; and Zha, H. 2017.
\newblock Fake news mitigation via point process based intervention.
\newblock In \emph{ICML}.

\bibitem[{Gervini and Gasser(2004)}]{gervini2004self}
Gervini, D.; and Gasser, T. 2004.
\newblock Self-modelling warping functions.
\newblock \emph{Journal of the Royal Statistical Society: Series B (Statistical
  Methodology)}, 66(4): 959--971.

\bibitem[{Gionis et~al.(1999)Gionis, Indyk, and Motwani}]{GionisIM1999hash}
Gionis, A.; Indyk, P.; and Motwani, R. 1999.
\newblock Similarity Search in High Dimensions via Hashing.
\newblock In \emph{VLDB}.

\bibitem[{Gogolou et~al.(2020)Gogolou, Tsandilas, Echihabi, Bezerianos, and
  Palpanas}]{gogolou2020data}
Gogolou, A.; Tsandilas, T.; Echihabi, K.; Bezerianos, A.; and Palpanas, T.
  2020.
\newblock Data series progressive similarity search with probabilistic quality
  guarantees.
\newblock In \emph{SIGMOD}.

\bibitem[{Guo et~al.(2018)Guo, Li, and Liu}]{initiator}
Guo, R.; Li, J.; and Liu, H. 2018.
\newblock INITIATOR: Noise-contrastive Estimation for Marked Temporal Point
  Process.
\newblock In \emph{IJCAI}.

\bibitem[{Gupta and Bedathur(2021)}]{gupta2021reformd}
Gupta, V.; and Bedathur, S. 2021.
\newblock Region Invariant Normalizing Flows for Mobility Transfer.
\newblock In \emph{CIKM}.

\bibitem[{Gupta et~al.(2021{\natexlab{a}})Gupta, Bedathur, Bhattacharya, and
  De}]{gupta2021learning}
Gupta, V.; Bedathur, S.; Bhattacharya, S.; and De, A. 2021{\natexlab{a}}.
\newblock Learning Temporal Point Processes with Intermittent Observations.
\newblock In \emph{AISTATS}.

\bibitem[{Gupta et~al.(2021{\natexlab{b}})Gupta, Bedathur, and De}]{tppsel}
Gupta, V.; Bedathur, S.; and De, A. 2021{\natexlab{b}}.
\newblock Supplementary Material for Learning Temporal Point Processes for
  Efficient Retrieval of Continuous Time Event Sequences.
\newblock \url{https://github.com/data-iitd/neuroseqret/}.

\bibitem[{Jaakkola et~al.(1999)Jaakkola, Haussler et~al.}]{fisher}
Jaakkola, T.~S.; Haussler, D.; et~al. 1999.
\newblock Exploiting generative models in discriminative classifiers.
\newblock In \emph{NeurIPS}.

\bibitem[{Jing and Smola(2017)}]{jing2017neural}
Jing, H.; and Smola, A.~J. 2017.
\newblock Neural survival recommender.
\newblock In \emph{WSDM}.

\bibitem[{Joachims(2002)}]{Joachims2002ranksvm}
Joachims, T. 2002.
\newblock Optimizing Search Engines Using Clickthrough Data.
\newblock 133--142. ACM.

\bibitem[{Kang and McAuley(2018)}]{sasrec}
Kang, W.-C.; and McAuley, J. 2018.
\newblock Self-Attentive Sequential Recommendation.
\newblock In \emph{ICDM}.

\bibitem[{Kumar et~al.(2019)Kumar, Zhang, and Leskovec}]{srijan}
Kumar, S.; Zhang, X.; and Leskovec, J. 2019.
\newblock Predicting dynamic embedding trajectory in temporal interaction
  networks.
\newblock In \emph{KDD}.

\bibitem[{Li et~al.(2020)Li, Wang, and McAuley}]{tisasrec}
Li, J.; Wang, Y.; and McAuley, J. 2020.
\newblock Time Interval Aware Self-Attention for Sequential Recommendation.
\newblock In \emph{WSDM}.

\bibitem[{Likhyani et~al.(2020)Likhyani, Gupta, Srijith, Deepak, and
  Bedathur}]{likhyani2020colab}
Likhyani, A.; Gupta, V.; Srijith, P.; Deepak, P.; and Bedathur, S. 2020.
\newblock Modeling Implicit Communities from Geo-tagged Event Traces using
  Spatio-Temporal Point Processes.
\newblock In \emph{WISE}.

\bibitem[{Liu et~al.(2014)Liu, Mu, Kumar, and Chang}]{liu2014discrete}
Liu, W.; Mu, C.; Kumar, S.; and Chang, S.-F. 2014.
\newblock Discrete graph hashing.

\bibitem[{Mei and Eisner(2017)}]{MeiE16}
Mei, H.; and Eisner, J.~M. 2017.
\newblock The neural hawkes process: A neurally self-modulating multivariate
  point process.
\newblock In \emph{NeurIPS}.

\bibitem[{Mei et~al.(2019)Mei, Qin, and Eisner}]{mei_icml}
Mei, H.; Qin, G.; and Eisner, J. 2019.
\newblock Imputing Missing Events in Continuous-Time Event Streams.
\newblock In \emph{ICML}.

\bibitem[{Mueen and Keogh(2016)}]{mueen2016extracting}
Mueen, A.; and Keogh, E. 2016.
\newblock Extracting optimal performance from dynamic time warping.
\newblock In \emph{KDD}.

\bibitem[{Mueen et~al.(2017)Mueen, Zhu, Yeh, Kamgar, Viswanathan, Gupta, and
  Keogh}]{mass}
Mueen, A.; Zhu, Y.; Yeh, M.; Kamgar, K.; Viswanathan, K.; Gupta, C.; and Keogh,
  E. 2017.
\newblock The Fastest Similarity Search Algorithm for Time Series Subsequences
  under Euclidean Distance.
\newblock Http://www.cs.unm.edu/~mueen/FastestSimilaritySearch.html.

\bibitem[{M{\"u}ller(2007)}]{muller2007}
M{\"u}ller, M. 2007.
\newblock Dynamic time warping.
\newblock \emph{Information retrieval for music and motion}, 69--84.

\bibitem[{Paparrizos and Gravano(2015)}]{paparrizos2015k}
Paparrizos, J.; and Gravano, L. 2015.
\newblock k-shape: Efficient and accurate clustering of time series.
\newblock In \emph{SIGMOD}.

\bibitem[{Qin et~al.(2020)Qin, Bai, and Sun}]{qin2020ghashing}
Qin, Z.; Bai, Y.; and Sun, Y. 2020.
\newblock GHashing: Semantic Graph Hashing for Approximate Similarity Search in
  Graph Databases.
\newblock In \emph{KDD}.

\bibitem[{Rakthanmanon et~al.(2012)Rakthanmanon, Campana, Mueen, Batista,
  Westover, Zhu, Zakaria, and Keogh}]{ucrdtw}
Rakthanmanon, T.; Campana, B.; Mueen, A.; Batista, G.; Westover, B.; Zhu, Q.;
  Zakaria, J.; and Keogh, E. 2012.
\newblock Searching and Mining Trillions of Time Series Subsequences under
  Dynamic Time Warping.
\newblock In \emph{KDD}.

\bibitem[{Rizoiu et~al.(2017)Rizoiu, Xie, Sanner, Cebrian, Yu, and
  Van~Hentenryck}]{rizoiu2017expecting}
Rizoiu, M.-A.; Xie, L.; Sanner, S.; Cebrian, M.; Yu, H.; and Van~Hentenryck, P.
  2017.
\newblock Expecting to be hip: Hawkes intensity processes for social media
  popularity.
\newblock In \emph{WWW}.

\bibitem[{Roy et~al.(2020)Roy, De, and Chakrabarti}]{roy2020adversarial}
Roy, I.; De, A.; and Chakrabarti, S. 2020.
\newblock Adversarial Permutation Guided Node Representations for Link
  Prediction.
\newblock \emph{arXiv preprint arXiv:2012.08974}.

\bibitem[{Saha et~al.(2018)Saha, Samanta, Ganguly, and De}]{saha2018crpp}
Saha, A.; Samanta, B.; Ganguly, N.; and De, A. 2018.
\newblock Crpp: Competing recurrent point process for modeling visibility
  dynamics in information diffusion.
\newblock In \emph{CIKM}.

\bibitem[{Salakhutdinov and Hinton(2009)}]{salakhutdinov2009semantic}
Salakhutdinov, R.; and Hinton, G. 2009.
\newblock Semantic hashing.
\newblock In \emph{International Journal of Approximate Reasoning}.

\bibitem[{Samanta et~al.(2017)Samanta, De, Chakraborty, and
  Ganguly}]{samanta2017lmpp}
Samanta, B.; De, A.; Chakraborty, A.; and Ganguly, N. 2017.
\newblock LMPP: a large margin point process combining reinforcement and
  competition for modeling hashtag popularity.
\newblock In \emph{IJCAI}.

\bibitem[{Sewell(2011)}]{sewell2011fisher}
Sewell, M. 2011.
\newblock The fisher kernel: a brief review.
\newblock \emph{RN}, 11(06): 06.

\bibitem[{Shchur et~al.(2020)Shchur, Bilo{\v{s}}, and
  G{\"u}nnemann}]{shchur2019intensity}
Shchur, O.; Bilo{\v{s}}, M.; and G{\"u}nnemann, S. 2020.
\newblock Intensity-Free Learning of Temporal Point Processes.
\newblock In \emph{ICLR}.

\bibitem[{Shelton et~al.(2018)Shelton, Qin, and Shetty}]{shelton}
Shelton, C.~R.; Qin, Z.; and Shetty, C. 2018.
\newblock Hawkes Process Inference with Missing Data.
\newblock In \emph{AAAI}.

\bibitem[{Shen et~al.(2018)Shen, Chen, Keogh, and Jin}]{shen2018accelerating}
Shen, Y.; Chen, Y.; Keogh, E.; and Jin, H. 2018.
\newblock Accelerating time series searching with large uniform scaling.
\newblock In \emph{SDM}.

\bibitem[{Su et~al.(2020)Su, Liu, Zheng, Zhou, and Zheng}]{su2020survey}
Su, H.; Liu, S.; Zheng, B.; Zhou, X.; and Zheng, K. 2020.
\newblock A survey of trajectory distance measures and performance evaluation.
\newblock \emph{VLDB Journal}.

\bibitem[{Tabibian et~al.(2019)Tabibian, Upadhyay, De, Zarezade, Sch{\"o}lkopf,
  and Gomez-Rodriguez}]{tabibian2019enhancing}
Tabibian, B.; Upadhyay, U.; De, A.; Zarezade, A.; Sch{\"o}lkopf, B.; and
  Gomez-Rodriguez, M. 2019.
\newblock Enhancing human learning via spaced repetition optimization.
\newblock \emph{PNAS}.

\bibitem[{Valera et~al.(2014)Valera, Gomez-Rodriguez, and Gummadi}]{Valera2014}
Valera, I.; Gomez-Rodriguez, M.; and Gummadi, K. 2014.
\newblock Modeling Diffusion of Competing Products and Conventions in Social
  Media.
\newblock \emph{arXiv preprint arXiv:1406.0516}.

\bibitem[{Vaswani et~al.(2017)Vaswani, Shazeer, Parmar, Uszkoreit, Jones,
  Gomez, Kaiser, and Polosukhin}]{vaswani2017}
Vaswani, A.; Shazeer, N.; Parmar, N.; Uszkoreit, J.; Jones, L.; Gomez, A.~N.;
  Kaiser, L.; and Polosukhin, I. 2017.
\newblock Attention is All You Need.
\newblock In \emph{NeurIPS}.

\bibitem[{Wang et~al.(2017)Wang, Fu, Liu, Hu, and Aggarwal}]{wang2017human}
Wang, P.; Fu, Y.; Liu, G.; Hu, W.; and Aggarwal, C. 2017.
\newblock Human Mobility Synchronization and Trip Purpose Detection with
  Mixture of Hawkes Processes.
\newblock In \emph{KDD}.

\bibitem[{Wehenkel and Louppe(2019)}]{umnn}
Wehenkel, A.; and Louppe, G. 2019.
\newblock Unconstrained monotonic neural networks.
\newblock In \emph{NeurIPS}.

\bibitem[{Xiao et~al.(2017)Xiao, Farajtabar, Ye, Yan, Song, and
  Zha}]{xiao2017wasserstein}
Xiao, S.; Farajtabar, M.; Ye, X.; Yan, J.; Song, L.; and Zha, H. 2017.
\newblock Wasserstein Learning of Deep Generative Point Process Models.
\newblock In \emph{NeurIPS}.

\bibitem[{Xu et~al.(2018)Xu, Carin, and Zha}]{ido}
Xu, H.; Carin, L.; and Zha, H. 2018.
\newblock Learning registered point processes from idiosyncratic observations.
\newblock In \emph{ICML}.

\bibitem[{Yoon et~al.(2019)Yoon, Jarrett, and van~der Schaar}]{yoon2019time}
Yoon, J.; Jarrett, D.; and van~der Schaar, M. 2019.
\newblock Time-series generative adversarial networks.
\newblock In \emph{NeurIPS}.

\bibitem[{Zamani~Dadaneh et~al.(2020)Zamani~Dadaneh, Boluki, Yin, Zhou, and
  Qian}]{dadaneh20a}
Zamani~Dadaneh, S.; Boluki, S.; Yin, M.; Zhou, M.; and Qian, X. 2020.
\newblock Pairwise Supervised Hashing with Bernoulli Variational Auto-Encoder
  and Self-Control Gradient Estimator.
\newblock In \emph{UAI}.

\bibitem[{Zhang et~al.(2010)Zhang, Wang, Cai, and Lu}]{zhang2010self}
Zhang, D.; Wang, J.; Cai, D.; and Lu, J. 2010.
\newblock Self-taught hashing for fast similarity search.
\newblock In \emph{SIGIR}.

\bibitem[{Zhang et~al.(2021)Zhang, Iyer, Tendulkar, Aggarwal, and
  De}]{zhang2021learning}
Zhang, P.; Iyer, R.; Tendulkar, A.; Aggarwal, G.; and De, A. 2021.
\newblock Learning to Select Exogenous Events for Marked Temporal Point
  Process.
\newblock In \emph{NeurIPS}.

\bibitem[{Zhang et~al.(2020)Zhang, Lipani, Kirnap, and Yilmaz}]{zhang2019self}
Zhang, Q.; Lipani, A.; Kirnap, O.; and Yilmaz, E. 2020.
\newblock Self-attentive Hawkes processes.
\newblock \emph{ICML}.

\bibitem[{Zuo et~al.(2020)Zuo, Jiang, Li, Zhao, and Zha}]{zuo2020transformer}
Zuo, S.; Jiang, H.; Li, Z.; Zhao, T.; and Zha, H. 2020.
\newblock Transformer Hawkes Process.
\newblock In \emph{ICML}.

\end{thebibliography}
%
\end{document}